\documentstyle[12pt,titlepage]{article}
\input epsf

\font\small=cmr8 scaled \magstep0

\outer\def\beginsection#1\par{\medbreak\bigskip
      \message{#1}\leftline{\bf#1}\nobreak\medskip
\vskip-\parskip
      \noindent}

\def\laq{\raise 0.4ex\hbox{$<$}\kern -0.8em\lower 0.62
ex\hbox{$\sim$}}
\def\gaq{\raise 0.4ex\hbox{$>$}\kern -0.7em\lower 0.62
ex\hbox{$\sim$}}

\def\beq{\begin{equation}}
\def\eeq{\end{equation}}
\def\bea{\begin{eqnarray}}
\def\eea{\end{eqnarray}}

\def \da {\delta}

\def \om {\omega}
\def \Om {\Omega}
\def \og {\Om_G}
\def \rg {\rho_G}

\begin{document}
\bibliographystyle {unsrt}

\titlepage
\begin{flushright}
CERN-TH/96-37 \\
BGU-PH-96/05 \\
\end{flushright}
\vspace{22mm}
\begin{center}
{\bf {\Large Peak and End Point of the Relic Graviton Background}\\
\bigskip
{\Large in String Cosmology}}\\

\vspace{10mm}
\centerline{R. Brustein$^{(a)}$, M. Gasperini$^{(b,c)}$
 and G. Veneziano$^{(c)}$}
\bigskip
\centerline{$^{(a)}${\it Department of Physics, Ben-Gurion
University,
Beer-Sheva 84105, Israel}}
\smallskip
\centerline{$^{(b)}${\it Dipartimento di Fisica Teorica,
Turin University, Via P.Giuria 1, 10125 Turin, Italy}}
\smallskip
\centerline{$^{(c)}${\it Theory Division, CERN, CH-1211, Geneva 23,
Switzerland} }
\vskip 2  cm

{\large  Abstract} 

\end{center}

\noindent
Using general arguments we determine the 
allowed region for the end point frequency and the 
peak energy density of the stochastic background 
of gravity waves expected in string cosmology. We provide
an accurate estimate of the minimal experimental sensitivity
required to detect a signal in the Hz to GHz range.

\vspace{5mm}

\vfill
\begin{flushleft}
CERN-TH/96-37\\
April 1996 

\end{flushleft}

\newpage

In a recent paper \cite{1} we computed, 
in collaboration with
M. Giovannini, the spectrum of relic gravity waves produced in the
context of the 
so-called ``pre-big-bang" scenario of string cosmology \cite{2,3}. 
 We showed that the
spectral energy density of the produced gravitons grows 
with frequency following 
a  Rayleigh-Jeans-type behaviour at low frequencies and then, after a 
possible flatter intermediate region,  reaches a peak
value $\Om_G(\om_1)\sim 10^{-5}$ (in critical units) at
$\om_1\sim10^2 ~{\rm GHz}$.  The stochastic background of
relic gravity waves is thus expected to be much stronger, at
high frequency, than in the context of the standard
inflationary scenario, which predicts, in the most favourable case, 
a flat spectrum  at a 
level \cite{4} of $\og \sim 10^{-14}$ in its higher 
frequency range.  Such an enhanced
production of high-frequency gravitons 
represents a typical  signature of the pre-big-bang 
scenario, as previously stressed in a number of papers
\cite{3,5,6}.

The explicit computation of the spectrum performed in \cite{1}
made use of a two-parameter model of
the metric--dilaton background  and of the equation for tensor
perturbations obtained from the low-energy string effective  
action.
 Such an equation may be questioned when
applied to the truly ``stringy" high-curvature regime 
in which all higher orders in the string
tension have to be
taken into account. In view of this,  the present
paper aims at confirming the main findings of \cite{1} by  
determining, within some inherent uncertainty, the position
and height of the peak signal from the expected graviton
background, without using either the perturbation equation
or an explicit parametrization of the shape of the spectrum. 
We also discuss to what extent the 
position and height of the peak are 
affected by late  entropy production, 
associated with some additional reheating process occurring well
below the string scale.

We shall work in the context of a
scenario \cite{1,2,3}, in which the Universe evolves from the
string perturbative vacuum, through a dilaton-driven phase
and a high-curvature stringy phase, towards the final
radiation-dominated epoch. For a detailed discussion of 
the initial, pre-big-bang epoch 
we refer the reader to more specific papers on the
general picture \cite{3,7}, on the underlying symmetries \cite{2,8},
on the perturbation spectra \cite{5,9} and on the difficulties of a
classical matching to the standard radiation era \cite{10}. The
main aspect of the  scenario that we shall use
here  is the fact
that the time evolution of the classical background 
amplifies, with similar efficiency, both metric perturbations 
(gravity waves) 
and the vacuum fluctuations of the electromagnetic \cite{11} 
and of other gauge 
fields, as a consequence of their coupling to a dynamical dilaton. 
 
Thus, unlike ordinary inflation, string cosmology
 naturally leads to a democratic  production of  
all sorts of ultra-relativistic particles \cite{12}, most of which 
subsequently  thermalize  and start 
dominating the energy density. Only gravitationally coupled 
particles, such as  gravitons and dilatons, 
 drop out of thermal equilibrium soon after the string phase.
Of course, such a thermal
background may possibly represent only a small fraction of the Cosmic
Microwave Background (CMB) that we now observe, if 
later, efficient sources of thermal entropy existed. Nevertheless,
because of their common origin at the same (string) scale,  the
energy density of the produced gravitons remains linked to the
energy density of this primordial thermal radiation \cite{12a},
and this link allows us to relate the peak of the graviton spectrum to
the present CMB temperature, $T_0= 2.7 ~{\rm K}$.

We start by recalling that, in our
scenario,  metric fluctuations are amplified with a spectrum 
that grows with
frequency. However, without knowing explicitly the time evolution of
the model during the string phase,  we cannot
compute exactly the maximal amplified
proper frequency $\om_1$. We thus {\it define} $\om_1$ as the
frequency corresponding to the production of one graviton per
polarization and per unit phase-space volume. It is known that,
for larger frequencies, the production has to be exponentially
suppressed \cite{13}. With this definition, the ``end point" of
the spectrum in the plane ($\om, \rg (\om)$), where $\rg
(\om)=d\rg /d\ln \om$ is the spectral energy density, has
coordinates $\om_1$ and $\rg(\om_1)=\om_1^4/\pi^2$.

We shall now relate these coordinates to the present
temperature $T_0$, and to the temperature scale $T_r$
marking the beginning of the phase dominated by thermal radiation, 
soon after the string era. 
Such a scale is defined by the Einstein equations as
\beq
H_r^2={8\pi \over 3M_p^2}{\pi^2 N_r\over 30} T_r^4
\label{1}
\eeq
where $M_p$ is the Planck mass, $H_r$ the Hubble factor at
$t=t_r$, and $N_r$ is the total effective number of massless
degrees of freedom in thermal equilibrium \cite{15} at $t=t_r$ 
(as $N_r \gg 1$, the graviton contribution to this equation is 
negligible). 
Let us also define the fraction $\da s$ of the present thermal
entropy density, generated at some intermediate scale
between $t_r$ and the present time $t_0$, as $\da
s=(s_0-s_r)/s_0$, where \cite{15}
\beq
s_0\equiv {2\pi^2\over45}n_0(a_0T_0)^3= {2\pi^2\over45}
n_r(a_rT_r)^3+s_0\da s \equiv s_r+s_0\da s~ .
\label{2}
\eeq
Here $n_0,n_r$ are the number of species contributing (each
with its own weight) to the thermal entropy at $t_0$ and
$t_r$, respectively, and $a_0, a_r$ are the corresponding scale
factors. By expressing $\om_1(t_0)$ as 
$\om_1(t_r) a_r/a_0$, 
and using eqs. (\ref{1}) and (\ref{2}), the present
coordinates of the end point of the spectrum can be written in
the form
\beq
\om_1(t_0)= T_0\left[M_s(t_r)\over M_p\right]^{1/2}
\left (8\pi^3N_r\over 90\right)^{1/4}
\left[{n_0\over n_r}(1-\da s)\right]^{1/3}
{\om_1(t_r)\over \sqrt{H_r M_s(t_r)}}
\label{3}
\eeq
\beq
\rg(\om_1,t_0)={\om_1^4(t_0)\over \pi^2}=\rho_\gamma
(t_0) \left[M_s(t_r)\over M_p\right]^2
{8\pi N_r\over 3N_0}
\left[{n_0\over n_r}(1-\da s)\right]^{4/3}
\left [ {\om_1(t_r)\over \sqrt{H_r M_s(t_r)}}
\right]^4~ .
\label{4}
\eeq
We have multiplied and divided by the value of the string
mass $M_s$ at the time $t=t_r$, and we have introduced the
present photon CMB energy density, $\rho_\gamma(t_0)=
(\pi^2 N_0/30)T_0^4$, where $N_0=2$ is the number of photon
degrees of freedom. Note that  eqs. (\ref{3}), (\ref{4}) are
exact, and that the time-dependence of  $M_s/M_p$ 
accounts for possible residual variations of the dilaton field
for $t>t_r$ (this time-dependence is attributed to $M_s$ or 
to $M_p$, depending on the frame in which one is working \cite{7}). 
Note also that $n_0, N_0$ are known
numbers of order unity, while $n_r, N_r$ are numbers of
order $10^2$--$10^3$, whose precise value depends on the superstring
model unifying gravity and gauge interactions.

We shall now discuss the uncertainty with which we can fix
the position of the peak of the spectrum in the plane ($\om,
\rg(\om)$), by using the two previous equations at fixed $\da
s$. We shall treat $\da s$ as a parameter that accounts for
all subsequent non-adiabatic processes, which are not
expected to be significant in our context, but which can in
principle dilute, to a certain extent, the primordial graviton
production (we assumed $\da s \ll 1$ in \cite{1}). We distinguish
two possibilities, which we shall discuss separately.

The first possibility, which seems to be favoured in our context,  
is the one in which $H_r\simeq M_s(t_r)\simeq \om_1(t_r)$. In this case, 
the total energy density $\rho_{qf}$ produced by the amplification of 
the vacuum fluctuations, which becomes critical at $t=t_r$, must satisfy 
\beq
{\rho_{qf}(t_r)\over M_s^4(t_r)}= {\pi^2 N_r \over 30}{T_r^4\over 
M_s^4(t_r)}={3\over 8\pi}{M_p^2\over M_s^2(t_r)}~ .
\label{4a}
\eeq
According to the above equation $\rho_{qf}$ cannot be much larger than 
$N_r M_s^4\simeq N_r \om_1^4$, otherwise $T_r$ would exceed $M_s$, which 
does not make sense in a string theory context. This implies that the 
integrated spectra are dominated by the end point values at $\om_1(t_r)
\simeq M_s(t_r)$. On the other hand, if $\rho_{qf}\simeq
	 N_r M_s^4$, the 
value of $M_p/M_s$ at $t=t_r$ is predicted from eq. (\ref{4a}) to be of 
order $N_r^{1/2}$, i.e. quite close to its present value. 
Therefore, for $H_r\simeq M_s(t_r)\simeq \om_1(t_r)$, the end point 
must coincide with the peak of the
spectrum, and the present position of the peak follows
directly from eqs. (\ref{3}) and (\ref{4}) with $M_s(t_r)$ fixed by a 
dilaton expectation value already in its present range (this is the case 
for which we computed an explicit spectrum \cite{1}).  

By inserting known numbers, and noting that $N_r\simeq
n_r$, we obtain in this case that for fixed $\da s$ the
peak position is controlled by the fundamental ratio 
$(M_s/M_p)$, whose present value is expected \cite{16} to lie in the
range $10^{-2}~\laq~ (M_s/M_p)~ \laq ~10^{-1}$. By using 
this range to define our uncertainty on the
peak position, we get
\beq
0.7\times 10^{11} {\rm Hz}\;(1-\da s)^{1/3}
\left(10^3\over n_r\right)^{1/12}
<~\om_1(t_0) ~ <
2\times 10^{11} {\rm Hz}~(1-\da s)^{1/3}
\left(10^3\over n_r\right)^{1/12}~.
 \label{5}
\eeq
This translates into an uncertainty for the height of the peak,
which can be written in units of critical energy density as
\beq
0.7\times 10^{-8} h_{100}^{-2}~(1-\da s)^{4/3}
\left(10^3\over n_r\right)^{1/3}
< ~\og (\om_1, t_0) ~<
0.7\times 10^{-6} h_{100}^{-2}~(1-\da s)^{4/3}
\left(10^3\over n_r\right)^{1/3}
 \label{6}
\eeq
(for the present CMB energy density, in critical units,  we
have used the value \cite{15} $\Om_\gamma (t_0)
=2.6\times 10^{-5}h_{100}^{-2}$, where $h_{100}=H_0(100~{\rm 
km~  sec^{-1}~ Mpc^{-1}})^{-1}$). 

The corresponding allowed region for
the peak of the spectrum is represented in {\bf Fig. 1} by two
boxes, which are obtained from eqs. (\ref{5}) and (\ref{6})
with $n_r=10^3$, for the two cases $\da s= 0$ and $\da
s=0.99$. Note that even if $99 \%$ of the present entropy was
produced during the latest stages of evolution, the graviton
signal stays well above the standard inflationary prediction,
which, in {\bf Fig. 1}, is represented by the flat spectrum
$\og=10^{-10}\Om_\gamma$.  We also note that the theoretical estimate 
for the maximal
allowed energy density, obtained from eq. (\ref{6}), is consistent 
with the bound obtained from nucleosynthesis, 
which implies, roughly, that the total energy density in gravitons 
cannot
exceed that of one massless degree of freedom in
thermal equilibrium. According to standard nucleosynthesis analysis 
 \cite{15,16a} we get in fact the bound \cite{16b} 
$\int \rho_G(\om, t_N) d\ln \om~ \laq ~0.1 \rho_R(t_N)$, where 
$\rho_R$ is the total radiation energy density at the freeze out of the 
neutron-to-proton ratio, $t=t_N$ (see however \cite{16c} for recent 
critical discussions of the standard nucleosynthesis analysis). When 
referred to the present CMB energy density, the above bound implies 
\beq
h^2_{100}\int \og (\om, t_0) d\ln \om~ < ~0.2 ~\Om_\gamma(t_0) 
h^2_{100} =0.5 \times 10^{-5} .
\label{nucleo}
\eeq
Unless $\om_1$ exactly coincides with the maximal allowed value of eq. 
(\ref{5}), the 
spectrum may even be flat, from the end point down to a
minimal frequency much smaller than one Hertz, 
without violating
such a bound. This situation is described by the dashed lines 
\cite{19a} of {\bf Fig. 1},
which define the allowed region for the maximal value of the spectral 
energy density, 
for the two cases $\da s=0$ and $\da s=99\%$.  

\centerline{\epsfxsize=4.0in\epsfbox{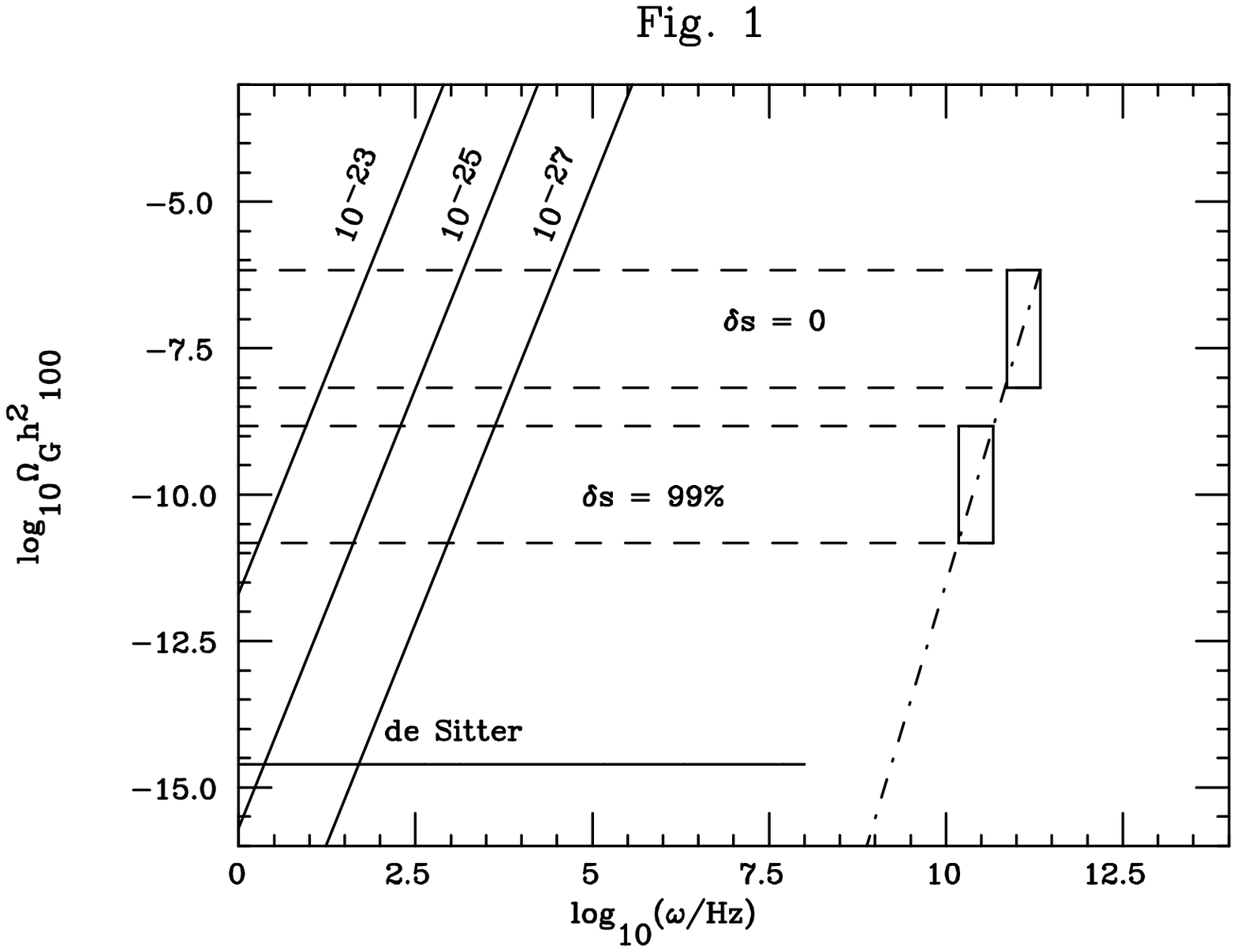}}

\noindent
\baselineskip=13 pt
{\small {  
{\bf Fig. 1}. {\em 
The area within the dashed lines 
defines the allowed region for the maximal value of the spectral 
energy density, 
for the two cases $\da s=0$ and $\da s = 0.99$  
(the plot is done using ${\rm n_r}=10^3$). The 
two boxes on the right border define the position of the end point of 
the spectrum if the end of the string era 
occurs in the strong coupling regime. For comparison, the flat
graviton spectrum of the de Sitter inflationary scenario is
plotted for an inflation scale high enough to account for the
observed large scale anisotropy. Also plotted are three lines of 
constant spectral amplitude $ S_h^{1/2}=10^{-23}, 10^{-25}$ and 
$10^{-27}    
{\rm Hz}^{-1/2}$, as well as the (dash-dotted) ``one-graviton" line, along 
which the end point is shifted as a function of late entropy 
production.}}}

\vskip 1 cm

\baselineskip=20pt
Let us now consider, for completeness, a scenario 
in which the  
curvature starts decreasing from the maximal scale
$H_1\simeq  
M_s(t_1)\simeq \om_1(t_1)$, while the string coupling 
$e^{\phi}$ ($\phi$ is the dilaton) is   
still very small. In this  scenario the transition to the regime of 
decelerated expansion is induced by 
higher derivative corrections rather than by
the back-reaction of the produced quanta. The
radiation-dominated epoch is now reached at a   
scale $H_r<<H_1$, and  is preceded by a decelerated,  
dilaton-driven
epoch \cite{7}. Inserting the explicit  
background solutions 
we find $\om_1(t_r)/\sqrt{H_r M_s(t_r)}<<1$ implying,  
from eq. (\ref{3}), that the end point of the spectrum is  
shifted to  much lower values  (unless $M_s(t_r)/M_p$ is very large;    
this seems to be excluded, however, since it
would correspond to the dilaton having gone very  
far into the
non-perturbative region at $t = t_r$). A shifted value of the 
end point $\om_1$ implies a 
smaller total energy density 
$\Omega_G$,  unless the spectrum
has  a peak at some arbitrary   
frequency $\om_P$ lower than 
$\om_1$, with such a height that the integrated graviton  
energy is still  of the same order as that of a thermal  
degree of freedom, at $t=t_r$. 
In that case the peak would  again be localized, for 
any given $\delta s$, within the dashed lines of {\bf Fig. 1}. The 
allowed region of {\bf Fig. 1} thus refers not only to a flat spectrum 
but also, in principle, to a spectrum with a peak energy density higher 
than the end point value. We note, however, that for 
$\om_P << \om_1$, and $\Omega_G(\om_P)>>\Omega_G(\om_1)$, 
present calculations based on the  
low-energy effective theory appear to preclude the possibility of 
having  enough  
quantum fluctuations to make them dominant 
 at $t=t_r$ (at least for a monotonic time evolution of the dilaton and 
of the metric scale factor). 

In order to compare our prediction with the sensitivities of 
gravity waves detectors, it is convenient
to express the spectral energy density in terms of the
spectral amplitude $S_h^{1/2}(\nu)$, $ \nu=\om/2\pi$, defined by
\beq
\langle h(\nu) h^\ast (-\nu')\rangle= {1\over 2}
\da(\nu+\nu')S_h(\nu) , ~~~~~~
h(\nu)=\int dt ~h(t) e^{-2\pi i\nu t} 
\label {7}
\eeq
where $h(x,t)$ is either one of the two polarized, 
dimensionless gravity wave amplitudes, and $\langle ... \rangle$  
denotes time or ensemble average. The average
energy density $\rg$, summing over polarizations, satisfies 
\cite{15} $8\pi \rg=M_p^2\langle\dot h^2\rangle$. The
corresponding spectral density, in critical units, is thus related
to $S_h$ by 
\beq
\og(\nu)={8\pi \rg(\nu) \over 3 M_p^2 H_0^2}= 
{4 \pi^2 \nu^3 S_h(\nu) \over 3 H_0^2}=
1.25\times
10^{36}h_{100}^{-2}~ \nu^3 S_h(\nu)~{\rm Hz}^{-2}. 
\label{8}
\eeq
In {\bf Fig. 1} we have plotted three lines of constant sensitivity, 
corresponding to $S_h^{1/2}=10^{-23}$, $10^{-25}$ and $10^{-27}$  
Hz$^{-1/2}$. 
Entering the region where we expect a signal, $\og
h_{100}^2~\laq ~10^{-6}$, would require a minimal 
sensitivity (from eq. (\ref{8}))
\beq
S_h^{1/2}(\nu)~\laq ~3 \times 10^{-26}\left({\rm
kHz}\over \nu \right)^{3/2}~{\rm Hz}^{-1/2}. 
\label{9}
\eeq

Very recent, direct measurements with cryogenic resonant
detectors provide an upper limit \cite{18} on the existence of
a relic graviton background, $S_h^{1/2}~ < ~6\times 10^{-22}~ 
{\rm Hz}^{-1/2}$, at $\nu=907$ Hz and $\nu=923$ Hz. This limit
is still too high to be significant for our background. However,
much better sensitivities can be reached through the
cross-correlation of existing resonant detectors \cite{18} such as
EXPLORER, NAUTILUS and AURIGA \cite{19}, as well 
as from interferometric 
detectors that will start 
operating in the near future,  such as GEO \cite{20}, LIGO
\cite{21} and VIRGO \cite{22}. Finally, spherical detectors \cite{23} 
also appear promising, because of their high
cross section at several frequencies for both tensor and scalar
metric fluctuations.

\vskip 1.5 cm
\noindent
\section*{Acknowledgements}
\noindent
R. B. is supported in part by an Alon Grant and by the Israel
Science Foundation. M. G. is supported in part by the CEE
contract ERBCHRX-CT94-0488. 
We thank B. Allen, P. Astone, V.
Braginski,  M. Cerdonio, E. Coccia, R. Drever, V. Fafone, M. Giovannini, 
G. Pallottino, E. Picasso, G. Pizzella, F. Ricci, K. Thorne and  S.
Vitale  for many useful discussions on the detection of a
stochastic gravity wave background. 
Special thanks are due to E. Kolb and E. Picasso for raising
stimulating questions about the validity of the graviton  
spectrum in string cosmology, and to G. Pizzella and the Rome
Group for a careful reading of a preliminary version of this
paper.


\newpage

\begin{thebibliography}{99}

\bibitem{1} R. Brustein, M. Gasperini, M. Giovannini
 and G. Veneziano, Phys. Lett. B361, 45 (1995).

\bibitem{2} G. Veneziano, Phys. Lett. B265, 287 (1991).

\bibitem{3}M. Gasperini and G. Veneziano, Astropart. Phys. 1, 317
(1993). 

\bibitem{4}L. M. Krauss and M. White, Phys. Rev. Lett. 69, 869 (1992).

\bibitem{5}M. Gasperini and M. Giovannini, Phys. Lett. B282, 36 
(1992);  Phys. Rev. D47, 1519 (1993).


\bibitem{6}M. Gasperini, in ``Proc. of the 2nd Journ\'ee 
Cosmologie" (Paris, June 1994), ed. by N. Sanchez and H. De
Vega (World Scientific, Singapore, 1995) p. 429.

\bibitem{7}
M. Gasperini and G. Veneziano, Mod. Phys. Lett. A8, 3701 (1993);
 Phys. Rev. D50, 2519 (1994).

\bibitem{8}
 K. A. Meissner and G. Veneziano, Phys. Lett. B267, 33 (1991); 
Mod. Phys. Lett. A6,  3397 
(1991); M. Gasperini and G. Veneziano, Phys. Lett. B277, 256 (1992); 
A. A. Tseytlin and C. Vafa, Nucl. Phys. B372, 443 (1992). 

\bibitem{9} R. Brustein, M. Gasperini, M. Giovannini,
V. Mukhanov and G. Veneziano, Phys. Rev. D51, 6744 (1995).

\bibitem{10} R. Brustein and G. Veneziano, Phys. Lett. B329, 429 
 (1994); N. Kaloper, R. Madden and K. Olive,
 Nucl. Phys. B452, 677 (1995); M. Gasperini, J. Maharana 
and G. Veneziano, 
{\em Graceful exit in quantum string cosmology}, 
Nucl. Phys. B (1996), in press (hep-th/9602087).

\bibitem{11}M. Gasperini, M. Giovannini and G. Veneziano, Phys.
Rev. Lett. 75, 3796 (1995); Phys. Rev. D52, 6651 (1995); 
D. Lemoine and M. Lemoine, ibid., 1955 (1995). 

\bibitem{12}G. Veneziano, {\em Status of string
cosmology: concepts and consequences}, in ``Proc. of the Int.
School of Astrophysics D. Chalonge" (Erice, September 1995), ed.
by N. Sanchez (Kluwer Acad. Pub., Dordrecht).  

\bibitem{12a}M. Gasperini, {\em Status of  
string
cosmology: phenomenological aspects}, in ``Proc. of the Int.
School of Astrophysics D. Chalonge" (Erice, September 1995), ed.
by N. Sanchez (Kluwer Acad. Pub., Dordrecht) (hep-th/9509127). 

\bibitem{13}L. Parker, Nature 261, 20 (1976); D. Birrel and P.
C. W. Davies, {\em  Quantum fields in curved space} (Cambridge U.
Press, Cambridge, England, 1982); B. Allen, Phys. Rev. D37, 2078
(1988).

\bibitem{15}See for instance 
E. W. Kolb and M. S. Turner, {\em The Early Universe} 
(Addison-Wesley P. Co., New York, 1990).

\bibitem{16} V. Kaplunovsky, Phys. Rev. Lett. 55, 1036 (1985). 

\bibitem{16a}T. Walker et al., Ap. J. 376, 51 (1991). 

\bibitem{16b}V. F. Schwartzmann, JEPT Lett. 9, 184 (1969). 

\bibitem{16c}N. Hata et al., Phys. Rev. Lett. 75, 3977 (1995); 
C. Copi et al., Phys. Rev. Lett. 75, 3981 (1995). 

\bibitem{19a}Strictly speaking, for the maximal value of the end point 
energy given in {\bf Fig. 1}, the nucleosynthesis bound might forbid a 
strictly flat spectrum down to the Hz scale. 
The effective 
deviation from flatness of the upper dashed line, required by the bound 
 (\ref{nucleo}), is however too small to be significant at the scale of 
{\bf Fig. 1}, and in the context of the approximations performed.

\bibitem{18}P. Astone et al., 
 {\em Upper limit for a gravitational wave stochastic background
measured  with the EXPLORER and NAUTILUS gravitational wave
resonant detectors} (Rome, February 1996), to appear.

\bibitem{19}
M. Cerdonio et al., {\em Status of the AURIGA Gravitational Wave
Antenna and Perspectives for the Gravitational Waves Search
with Ultracryogenic  Resonant Mass Detectors}, in ``Proc. 
of the First  Edoardo Amaldi Conference", ed.  by E. Coccia, G.
Pizzella and F. Ronga (World Scientific,  Singapore, 1995).

\bibitem{20} K. Danzmann et al.,  {\em GEO 600: A 600 m Laser
Interferometric 
  Gravitational Wave Antenna}, in ``Proceedings of the First 
Edoardo Amaldi Conference", ed by. E. Coccia, G. Pizzella and F.
Ronga  (World Scientific, Singapore, 1995).

\bibitem{21} A. Abramovici et al., Science 256, 325 (1992).

\bibitem{22} B. Caron et al., Nucl. Instrum. Meth. A360, 258
(1995).

\bibitem{23}E. Coccia, A. Lobo and J. Ortega, Phys. Rev. D52,
3735 (1995); see also W. W. Johnson and S. M. Merkowitz, Phys. Rev. 
Lett. 70, 2367 (1993); C. Z. Zhou and P. F. Michelson, Phys. Rev. D51, 
2517 (1995). 

\end{thebibliography}
\end{document}